\documentclass{aa}
\usepackage{graphicx}
\usepackage{txfonts}

\begin{document}

\title{Hickson 62. I. Kinematics of NGC4778.}
\author{Marilena Spavone\thanks{e-mail: spavone@na.infn.it}
          \inst{1}
          \and
          Enrichetta Iodice\inst{2}
          \and
          Giuseppe Longo\inst{1,}\inst{2,}\inst{3}
          \and\\
          Maurizio Paolillo\inst{1,}\inst{3}
          \and
          Silvia Sodani \inst{1}
          }

 \institute{Dipartimento di Scienze Fisiche, Universit\'a Federico II, via Cinthia 6, I-80126 Napoli, Italy
         \and
             INAF, Osservatorio Astronomico di Capodimonte, via Moiariello 16, I-80131 Napoli, Italy
         \and
             INFN, Sezione di Napoli, via Cinthia 6, Napoli, Italy
             }
\abstract
{Detailed studies of the photometric and kinematical properties of
compact groups of galaxies are crucial to understand the physics
of galaxy interactions and to shed light on some aspects of galaxy
formation and evolution.  In this paper we present a kinematical
and photometrical study of a member, NGC4778, of the nearest
(z=0.0137) compact group: Hickson 62.}
{The aim of this work was to investigate the existence of
kinematical anomalies in the brightest group member, NGC4778 in
order to constrain the dynamical status and the formation history
of the group. }
{We used long-slit spectra obtained with FORS1 at VLT, to measure
line-of-sight velocity distributions by means of the Fourier
Correlation Quotient method, and to derive the galaxy rotation curve
and velocity dispersion profile.  }
{Our analysis reveals that Hickson 62a, also known as NGC4778, is
an S0 galaxy with kinematical and morphological peculiarities,
both in its central regions (r $< 5''$) and in the outer halo. In
the central regions, the rotation curve shows the existence of a
kinematically decoupled stellar component, offset with respect to
the photometric center. In the outer halo we find an asymmetric
rotation curve and a velocity dispersion profile showing a rise on
the SW side, in direction of the galaxy NGC4776.}
{The nuclear counterrotation, the distorted kinematics in the
outer halo and the X-ray properties of the group suggest that
NGC4778 may be the product of a recent minor merger, more reliable
with a small late-type galaxy.}

    \authorrunning{Spavone et al. 2006}
  \titlerunning{Kinematics of NGC4778}
   \maketitle

\section{Introduction}\label{introduction}
Poor groups of galaxies are the most common cosmic structures and
contain a large fraction of the galaxies present in the universe
(\cite{Tully87}, \cite{Eke04a}). At a difference with rich clusters,
they span a wide range of densities, from loose groups, having spatial
density of baryonic matter slightly above that of the field, to
compact ones having densities comparable or higher than those
encountered in the cores of the richest clusters.  For this reason,
they are the ideal ground where to test all scenarios for galaxy
formation and evolution and where to pinpoint the details of the
physics controlling galaxy interactions.

Loose groups having masses in the range $10^{13} - 10^{14} \
M_{\odot}$, almost certainly are still collapsing and are
therefore crucial to uncover the formation processes shaping
cosmic structures (\cite{Zabludoff98}). Many factors converge in
identifying compact groups as good candidates to be one of the
regions where some of this processes occur. In first place, their
high spatial density of luminous matter and small velocity
dispersions imply dynamical lifetimes of the order of a fraction
of the Hubble time. This leading to the possibility that the
groups observed in the present time and in the local universe are
second generation objects, just accreting new members from the
loose groups of galaxies in which almost always they are embedded
(\cite{Vennik93}).  Second, compact groups are numerous and
contain a non negligible fraction of the baryonic matter in the
nearby universe (\cite{Pildis96}). Therefore, whatever is their
ultimate fate, they are bound to have an impact on the observable
properties of galaxies and cosmic structures.

As it was stressed in \cite{Mendes03} (hereafter M03), the
influence of environmental effects on the internal dynamics and
matter distribution of compact group galaxies has not yet been
clearly established, mostly due to lack of reliable kinematic
data. Extensive kinematical studies of both the stars and gas in
galaxies belonging to compact groups (\cite{Rubin91};
\cite{Nishiura00}; M03; \cite{Rampazzo98}; \cite{Bonfanti99}) all
suggest that peculiar kinematical behaviors are much more common
($\sim 75\%$) than in the field.

Moreover, M03 showed that velocity fields of the ionized gas
component in galaxies belonging to compact groups are often
significantly affected by non-circular motions, local asymmetries
and misalignments between the kinematic and stellar axes. These
peculiarities, however, tend to smooth out if the rotation curve
is derived by averaging the velocity fields of the galaxies over
large regions.  If these averaged values are used, a large
fraction ($\sim 80\%$) of the HCG members follow the same
Tully-Fisher (TF) relationship of field galaxies (M03). This may
indicate that the haloes of compact group galaxies have not been
significantly stripped inside their optical size. However,
according to M03, the remaining 20\% of the galaxies, including
the lowest-mass systems, present significant anomalies which could
be explained by assuming that compact group galaxies have smaller
dark halos than their field counterparts, due to tidal truncation.
A result which finds support in numerical simulations (cf.
\cite{Governato91}) and has important consequences on the groups
dynamical lifetimes.

In spite of the vast literature existing, due both to the limited
statistics and to the problems encountered in disentangling true
groups from optical ones as well as in deprojecting the measured
kinematic and photometric quantities, our understanding of the
dynamical and evolutionary status of compact groups still presents
quite a few gaps.  Gaps which can be filled only through detailed
multitechnique and multiwavelength analysis of individual cases.  In
this respect, the dynamical and evolutive status of a group two
observables are crucial: the detailed kinematics of the individual
galaxies and the structure of the diffuse hot gas halo.  In this paper
and in Paper II (\cite{Sodani06}) we present a study of the compact
group of galaxies Hickson 62 based on archival optical, spectroscopic
and X-ray data extracted from the ESO and the Chandra archives.

In this first paper we focus mainly on the peculiar kinematics and
on the photometry of the dominant galaxy NGC4778 (Hickson 62a),
while in Paper II we shall discuss the diffuse X ray halo
embedding at least two of the group components. This paper is
structured as follows: in Section \ref{HCG62} we discuss the main
characteristics of Hickson 62, in Section \ref{thedata} we
describe the observations and the data reduction procedure.  The
photometric properties of NGC4778 are presented in Section
\ref{phot} and the kinematics in \ref{kin}. Finally, we draw our
conclusions in \ref{discussion}. Throughout this paper we shall
adopt a distance of 60.9 Mpc based on $H_{0}= 70$ km
$\mbox{s}^{-1} \mbox{Mpc}^{-1}$ and an heliocentric radial
velocity $V = 4260$ km $\mbox{s}^{-1}$, this implies 1 arcsec =
0.29 kpc.

\section{HCG62}\label{HCG62}
Hickson~62 is a quartet of accordant early type galaxies at a
redshift of 0.0137 (Hickson, Kindl \& Hunchra, 1988). It also
belongs to the loose group LGG 313 (Rood \& Struble 1994;
Tovmassian 2001) containing at least 13 galaxies within a distance
of $\sim 200$ Kpc. The group is dominated by the pair NGC4778
(also known as HCG~62a) and NGC4776 (HCG~62b), having a projected
separation of only 8 Kpc. NGC4778 formerly classified as an E3
galaxy, is now classified as an S0 with a bright compact nucleus,
as already noticed by J. Herschel (1811), and subsequently
discovered to be a low luminosity AGN (\cite{Fukazawa01} and
\cite{Coziol98}). Both NGC4776 and NGC4761 (HCG62c) are classified
as peculiar S0's. Finally, NGC4764 (HCG62d) is a faint E2 galaxy
slightly more distant from the center of the action. The main
photometric and morphological characteristics of the group
galaxies are summarized in Table \ref{Hickson 62}.

The compact group is also embedded in a bright X-ray halo which
extends out to 200 kpc, revealing the presence of a deep common
gravitational well centered on NGC 4778. The high resolution
Chandra images also showed the presence of large cavities in the
gaseous halo due to the interaction of relativistic plasma with
the hot IGM, a sign of recent activity due to the NGC 4778 central
AGN (\cite{Vrtilek02}).

NGC4778 has been the target of several studies. Long slit
spectroscopy has been obtained at different position angles.
\cite{Bettoni95} and \cite{Rampazzo98}, positioned the slit along
the line joining the nuclei of NGC4778 and NGC4776
(P.A.=$127^{\circ}$) and their results lead to the conclusion that
the pair NGC4778/4776 is not interacting and the kinematical
peculiarities observed in NGC4778 are likely due to an interaction
with NGC4761. In fact, the velocity dispersion and rotation curves
of NGC4776 are well behaved and appear unperturbed, while the
velocity dispersion profile of NGC4778 shows a relatively sharp
increase to the SE, suggestive of the presence of a perturber.

\begin{table*}\label{Hickson 62}
\begin{tabular}{lcccc}
\hline
\hline
                  & NGC4778 (HCG62a)         &NGC4776 (HCG62b)         & NGC4761 (HCG62c)         & NGC4764 (HCG62d)\\
R.A.$_{J2000}$    & 12  53  06      &12  53  05      & 12  53  10      & 12  53  06.6\\
$\delta_{J2000}$  & -09  12  14     &-09  12  00     & -09  11  55     & -09 15  28\\
Type              & S0              &S0              & S0              & E2     \\
Size (arcsec)     & $52.3\times34.3$&$44.2\times33.3$&$24.6\times 12.3$&$19.9\times 16.8$\\
Redshift            & 0.0142          &0.01188         & 0.01478         &  0.01392\\
$m_v$             & 13.47             &14.04             & 14..86          &15.98\\
$m_b$             & 13.79             &14.21           & 15.00           &  16.30\\
$m_r$             & 11.25             &12.04             & 13.59           &    14.11\\
Ext.              & 0.227           &0.227           & 0.224           &    0.217\\
\hline
\end{tabular}

\caption{Summary of the properties of the Hickson 62 members. Data
are taken from Nasa/IPAC Extragalactic Database.}
\end{table*}
\section{Observations and data reduction}\label{thedata}
\subsection{Spectroscopic data}\label{spectroscopy}
\begin{figure*}
   \centering
   \includegraphics[width=10cm]{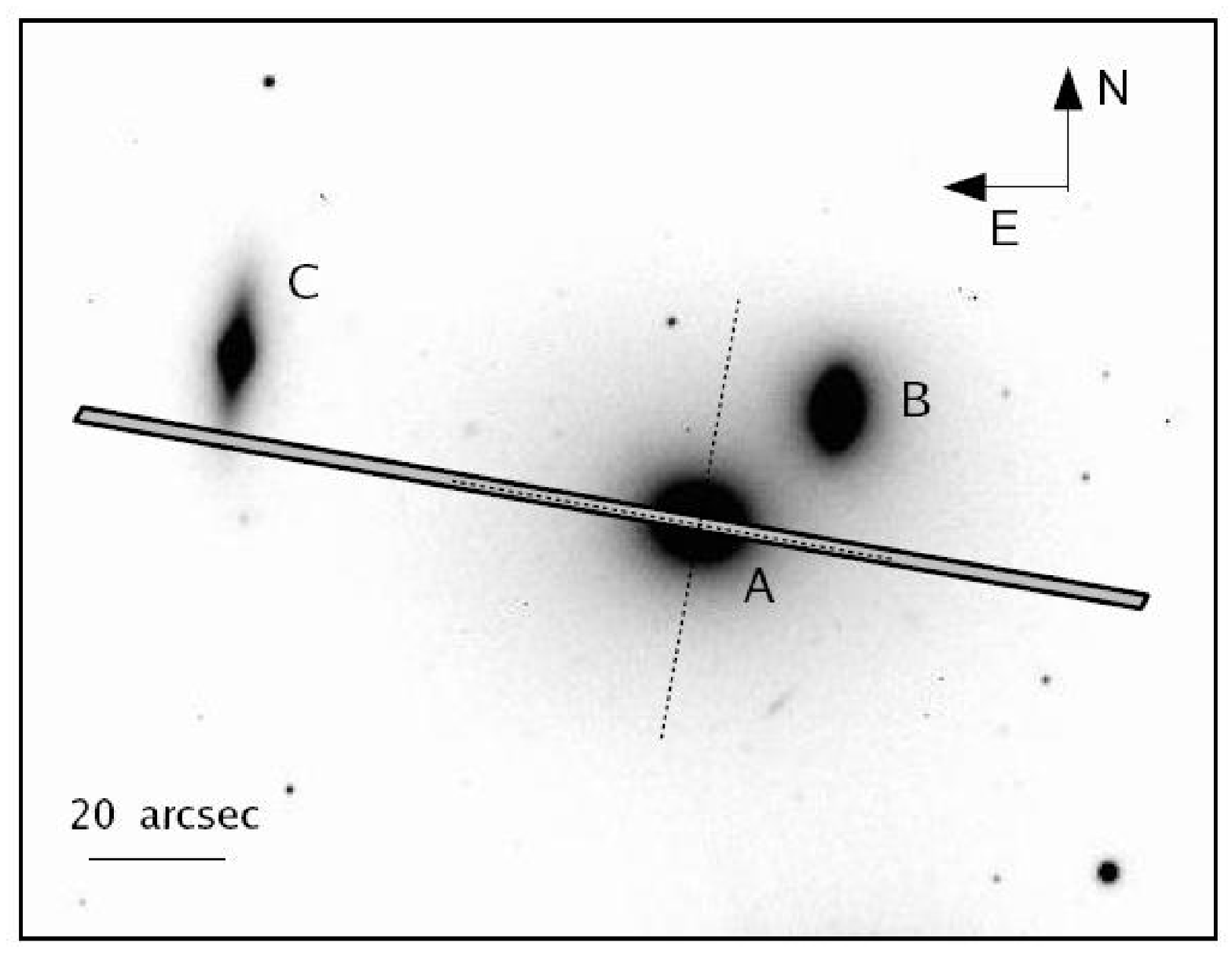}
      \caption{B band image of HCG62 with the slit overlayed.
The dashed lines overlayed on the slit represent the directions
along which we have extracted the light profiles presented in
fig.\ref{fit}.}
         \label{slit}
   \end{figure*}

The spectroscopic data were extracted from the European Southern
Observatory (ESO) public
archive\footnote{http://archive.eso.org/}.  They have been
obtained with the FORS1 spectrograph at VLT-UT1.  The detector is
a 2048 $\times$ 2048 pxl Tektronix CCD, with a scale of 0.2$''$
pixel$^{-1}$ (with the standard resolution collimator). The data,
consisting in four set of spectra, were acquired with a slit
1.6$''$ wide and $6'.8$ long, using the GRIS-600V grism with a
dispersion of 49 \AA\ mm$^{-1}$, corresponding to 1.18 \AA\
pxl$^{-1}$, in the $4650-7100$ \AA\ wavelength range.
The spectra were acquired along the photometric major axis
(PA=$80^{\circ}$) of NGC4778 and, by chance, they also intercepted
NGC4761 in a direction parallel to its minor axis, slightly
offcentered to SE side with respect to the nucleus (see
Fig.\ref{slit}).
The total integration time of the spectra is 2700 s and the average
seeing turned out to be 1$''$. A set of spectra of standard template
$F$ stars, were also acquired with the same configuration.

\begin{figure*}
   \centering
\includegraphics[width=8cm]{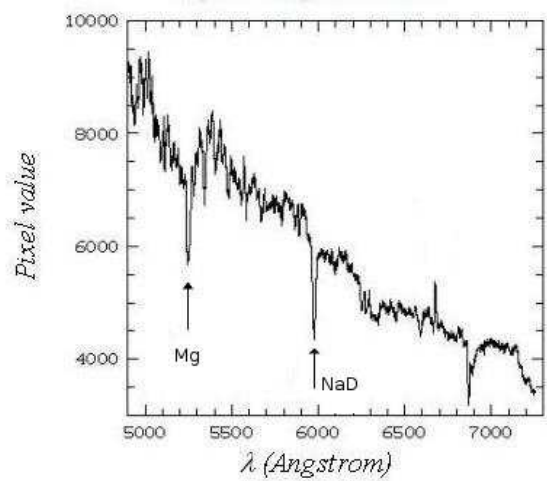}
\caption{Observed spectrum of NGC4778. The main absorption
features used for the kinematical analysis are marked.}
\label{spectrum} \end{figure*}

Individual frames were pre-reduced using the standard MIDAS image
processing package; the wavelength calibration
%
was made using the IRAF TWODSPEC.LONGSLIT package and a set of
He-Ar-Ne lamp spectra, taken for each observing night.  The
spectral resolution turned out to be 2.8 \AA\ (FWHM), equivalent
to a velocity resolution of $\sigma \simeq 60\ km s^{-1}$.  Sky
subtraction was performed using a narrow region at both edges of
the slit where there was minimum galaxy contamination. Finally,
all exposures were co-added in a final median averaged 2D spectrum
(cf. Fig.\ref{spectrum}).
The final steps of spectral processing consisted of \emph{i})
binning the spectra along the spatial direction in order to
achieve a signal-to-noise $(S/N)\geq 50$ at all radii (which is
the S/N measured at the last data points, while the central pixels
have S/N about 3 times larger), leaving no more than 2 data points
within the seeing disk; \emph{ii}) removing the galaxy continuum
by fitting a fourth order polynomial (for a detailed description
of the procedures see \cite{Bender94}).

\subsection{Imaging data for NGC4778}\label{photometry}

We also extracted from the ESO archive CCD images of Hickson 62
obtained with FORS1 at ESO VLT-UT1 in the Jonson B and R bands.
The CCD was the same used for the spectra, and the exposure times
were 240 seconds and 180 seconds in the B and R band,
respectively.  All images were taken in fairly good (i.e. FWHM
$\simeq 1''$) seeing conditions. The raw CCD frames were
pre-processed using the MIDAS image processing package and
standard techniques for de-biasing and flat-fielding.
Unfortunately, standard stars were not available to perform
absolute photometric calibration and we were forced to adopt the
tabulated zero point, namely: 27.594 \(\pm\) 0.045 and 27.939
\(\pm\) 0.025 for the B and R band, respectively.  We used the
IRAF-ELLIPSE task on the $B$ band image to perform the isophotal
analysis, to derive the effective parameters (the effective radius
turns out to be $R_e = 13\pm\ 1\ arcsec$), and to derive the
diskyness parameter, ellipticity and position angle for the
NGC4778 isophotes.

\section{Photometry of NGC4778}\label{phot}

Here we describe and discuss the main features in the light
distribution of NGC4778, in order to better understand the
kinematical properties described in section \ref{kin} and to
examine some possible connection between photometry and
kinematics.

\begin{figure*}
   \centering \includegraphics[width=10cm]{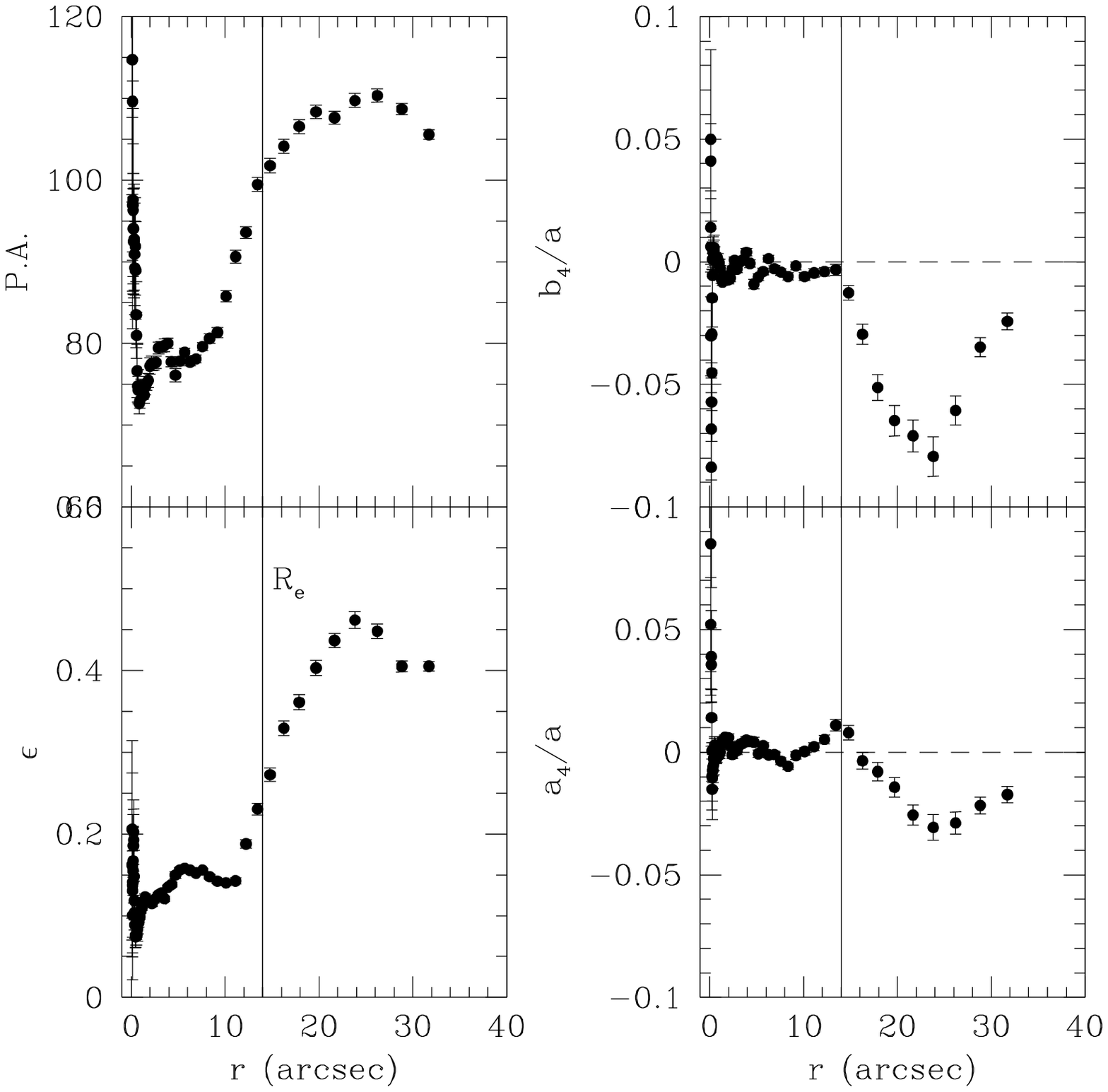}
   \includegraphics[width=10cm]{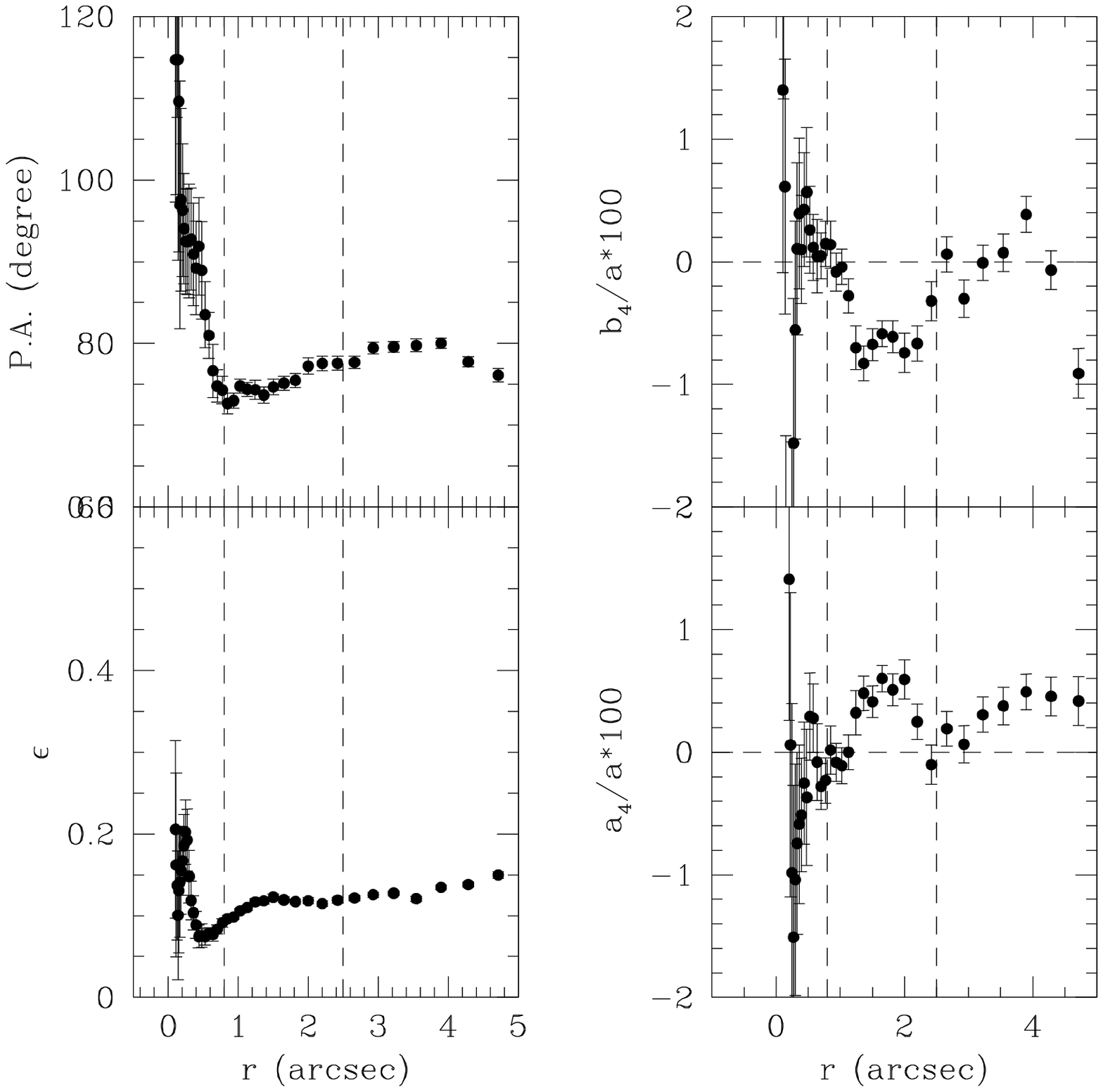} \caption{Top panel - Ellipticity,
   position angle (P.A.) and diskiness for NGC4778. The solid line indicates $R_{e}$. The trends of all parameters are different from those of $r<R_{e}$.
   Bottom panel - The same as above for the nuclear region of
   NGC4778. The first dashed line indicates the limit over which the data points are not affected by the seeing. The second dashed line marks the region discussed in the text.}\label{iso}
\end{figure*}

In Fig.\ref{iso} (top panel) we show the results of the isophotal
analysis of NGC4778. For $2'' \leq r \leq 12''$, the ellipticity
($\epsilon$), Position Angle (P.A.) and diskyness ($a_{4}/a$) are
approximately constant, thus indicating that in the intermediate
regions of the galaxy the isophotes are almost round, co-axial and
do not significantly deviate from purely elliptical shape
(\cite{Bender94}).
At larger radii (for $r \geq 12'' \sim R_e$), isophotes present
increasing flattening and become more boxy (i.e. $a_{4}/a < 0$).
Also, the P.A. changes of about $\sim 30^{\circ}$. These results
are consistent with those of \cite{Mendes92}. Looking in more
detail the nuclear regions (Fig.\ref{iso} bottom panel), between
$1''\leq r < 2.5''$ (outside the seeing disk), a small twisting
($\sim 10^{\circ}$) and an increasing flattening is observed. In
the same regions $a_{4}/a$ is significantly larger than zero.

\begin{figure*}
   \centering \includegraphics[width=10cm]{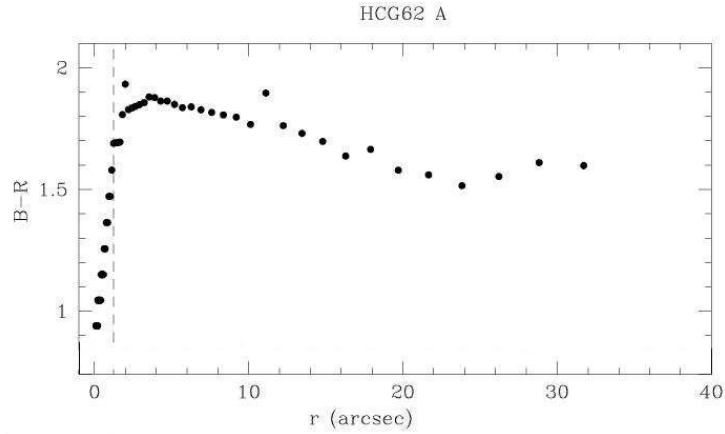}
   \caption{
   Mean B-R color profile of NGC4778. The dashed line indicates the limit of
reliability of the photometry.} \label{color}
\end{figure*}

In Fig.\ref{color} we show the average $B-R$ color profile of
NGC4778. The galaxy has bluer colors in the central regions and,
more precisely, we note an inversion of the trend of the color
profile beyond $5''$. Even though colors may be affected by
systematic errors due to the adopted calibration zero point, the
mean color profile of NGC4778 is consistent with the range of
values typical for early-type galaxies ($1.65 \leq B-R \leq 1.8$
\cite{Fukugita95}) and for spheroidal galaxies in compact groups
($0.9 \leq B-R \leq 1.7$ for the Es and $1.5 \leq B-R \leq 1.8$
for the S0; see \cite{Zepf91}).
\subsection{The model of the luminosity distribution of NGC4778}
We performed the 2-Dimensional model of the NGC4778 light
distribution through the super-position of a spheroidal central
component and an exponential disk (\cite{Iodice01}; Byun \&
Freeman 1995). The projected light of the spheroidal component
follows the generalized de Vaucouleurs law (\cite{Caon93}):
\begin{equation}
\mu_{b}(x,y)=\mu_{e}+k
\left[\left(\frac{r_{b}}{R_{e}}\right)^{1/n}-1 \right]
\end{equation}
with $k=2.17n-0.355$,
$r_{b}=\left[{x^{2}+y^{2}/q_b^{2}}\right]^{1/2}$, while $q_{b}$,
$\mu_{e}$ and $R_{e}$ are the {\it apparent axial ratio}, the {\it
effective surface brightness} and the {\it effective radius}
respectively. The projected light distribution of the exponential
disk (Freeman, 1970) is given by
\begin{equation}
\mu_{d}(x,y)=\mu_{0}+1.086\left(\frac{r_{d}}{r_{h}}\right)
\end{equation}
with $r_{d}=\left[{x^{2}+y^{2}/q_d^{2}}\right]^{1/2}$, and where
$q_{d}$ $\mu_{0}$ and $r_{h}$ are the {\it apparent axial ratio},
{\it the central surface brightness} and the {\it scalelength} of
the disk respectively.

We used: i) a model made by a bulge-like component with a Sersic
law $r^{1/n}$; ii) a model made by the superposition of a
bulge-like component with an $r^{1/n}$ light distribution, and an
exponential disk. To take into account the systematic effect of
the seeing that influence the model parameters, we masked the
central area of the galaxy (within the seeing).

The better agreement of the second model with the data, strongly
supports the conclusion that NGC4778 is a misclassified S0 galaxy.
The structural parametres derived from the fit are:
$\mu_{b}=23.48\pm0.01$, $r_{h}=21.9\pm0.2$arcsec,
$\mu_{e}=22.55\pm0.01$, $r_{e}=4.60\pm0.04$ and $n=1.49\pm0.01$,
leading to a bulge-to-disk ratio, $L_{B}/L_{D}\simeq0.24$.

\begin{figure*}
   \centering
   \includegraphics[width=10cm]{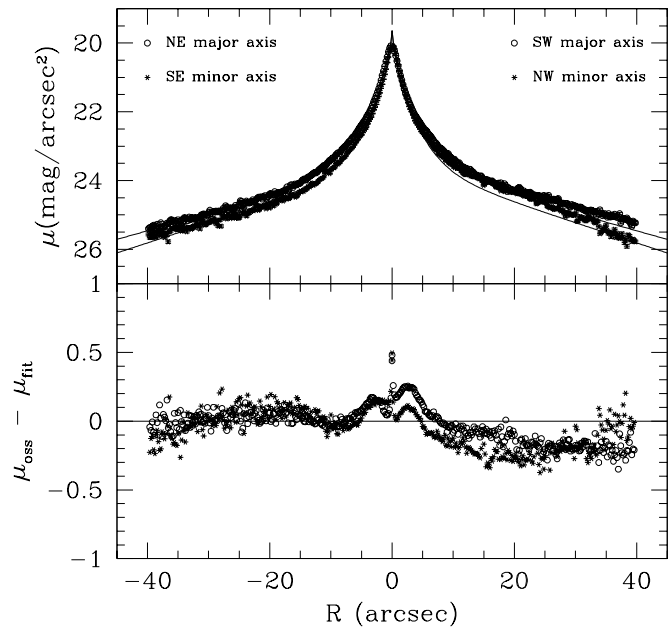}
   \caption{Top panel - 2-D fit of NGC4778 light distribution. The observed light profiles along the major (open circles), and minor axis (asterisk), are compared with those derived by the fit (continuous line).
   Bottom panel - Residuals between the observed and the fitted light profiles.}
\label{fit}
\end{figure*}

Fig.~\ref{fit} shows the comparison between the observed and
calculated light profiles in the B band. Residuals show that, in
the SE and NE directions the fitted profiles are in agreement with
the observed one. In the NW and SW directions instead, the
residuals show that the observed light profile is brighter than
the fitted one, due to light contamination from NGC4776. Notice
that the region between $1''$ and $5''$ from the center, the fit
does not reproduce the observed light profile thus suggesting the
presence of an additional component for which the light
distribution does not follow an $r^{1/n}$ law (\cite{Jog06}).

\section{\emph{Line-Of-Sight Velocity Distribution} of HCG62}\label{kin}

The \emph{Line-Of-Sight Velocity Distribution} (LOSVD) was then
derived from the continuum-removed spectra using the Fourier
Correlation Quotient (FCQ) method (\cite{Bender90};
\cite{Bender94}). We assumed the LOSVD to be described by a
Gaussian function and derived the line-of-sight rotational
velocity $v$ and the velocity dispersion $\sigma$. Main sources of
statistical and systematics errors are the template mismatching
and inaccurate continuum removal (the problem of error estimate
was analyzed in detail by \cite{Bender94}, \cite{Mehlert00} and
\cite{Gerhard93}). The errors on each kinematic data-point were
derived from photon statistics and CCD read-out noise and were
calibrated via Monte Carlo simulations: noise is added to the
template star, it is broadened according to the observed values of
$\sigma$, the output kinematical values were compared with the
input ones.

\begin{figure*}
   \centering \includegraphics[width=10cm]{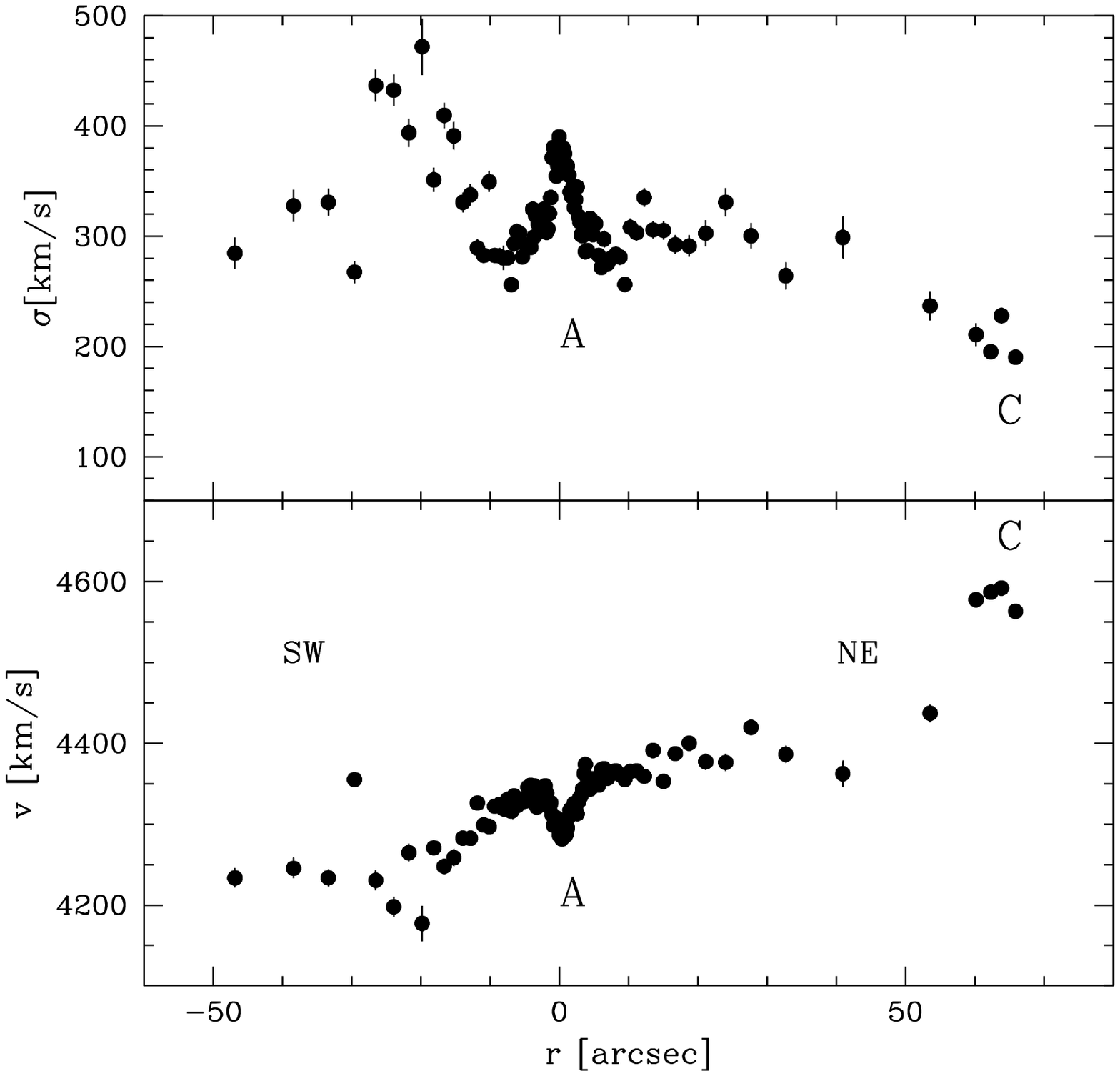}
   \includegraphics[width=10cm]{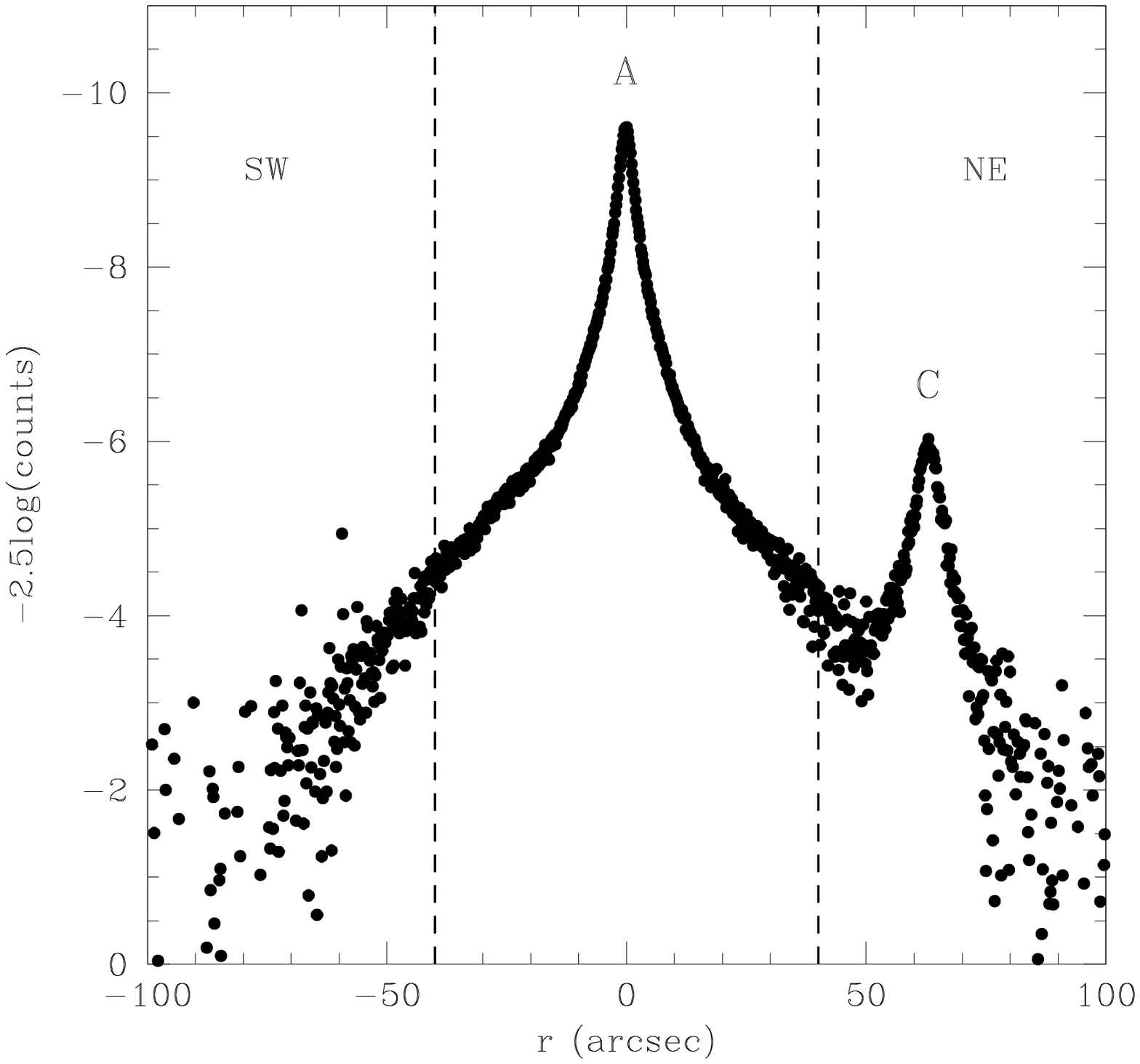}
\caption{Top panel -
   Rotation curve and velocity dispersion profile derived for the
   whole slit length, which include both galaxies NGC4778 and NGC4761 (labelled on the plot as A and C respectively). Bottom panel - Uncalibrated light profile through the slit,
   which includes both galaxies NGC4778 and NGC4761; the dashed lines
   indicates the region of the spectrum used to derive the
   kinematics.}
\label{tot}
\end{figure*}

In Fig.\ref{tot} (top panel) we show the line-of-sight radial
velocity curve and the velocity dispersions profile for both
NGC4778 and NGC4761. As the uncalibrated light profile through the
slit (see Fig.\ref{tot} bottom panel) also shows, the whole
kinematic profile extends up to about 60 arcsec ($\sim 17$ kpc)
from the galaxy center, which is about $4 R_{e}$ and the
contribution of NGC4761 to the whole spectrum is always about
$20\%$. For the study of NGC4778 we therefore extracted a region
in the spectra of 50 arcsec from the galaxy center.

\subsection{Kinematics of NGC4778}

\begin{figure*}
   \centering
   \includegraphics[width=10cm]{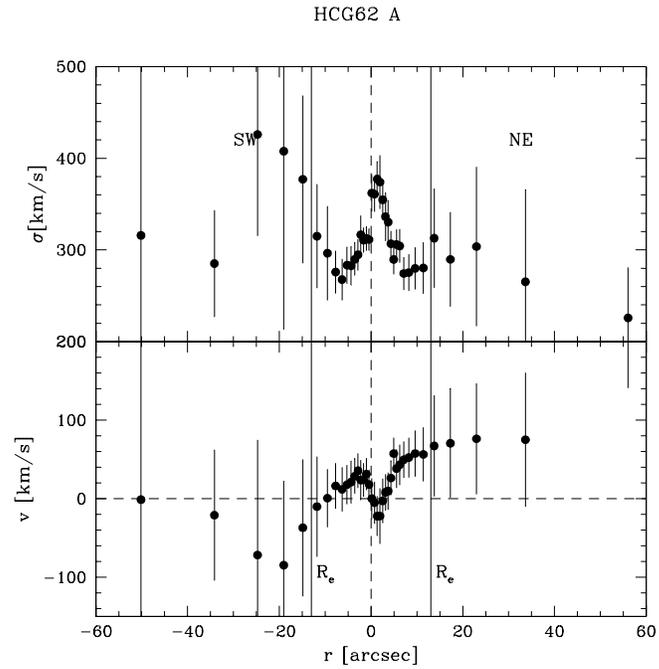}
   \includegraphics[width=10cm]{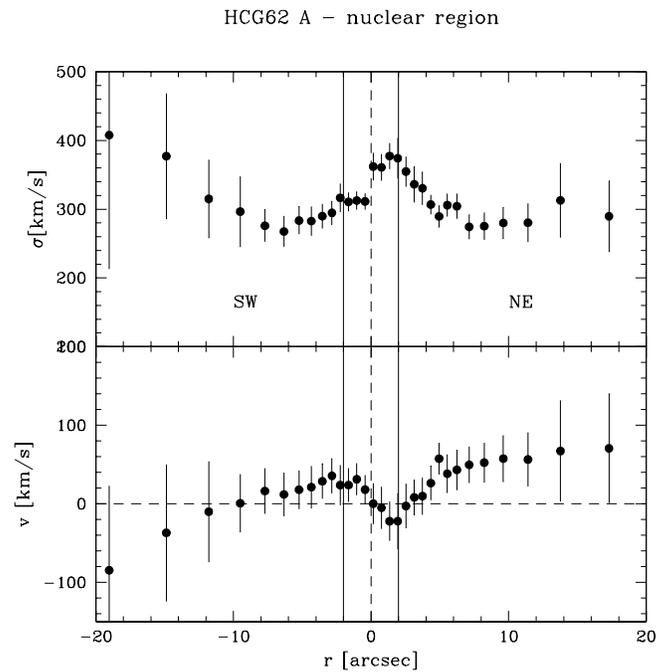}
\caption{Rotation curve and velocity dispersion profile derived
for NGC4778 for the whole
   galaxy extension (top panel) and for the nuclear regions (bottom
   panel) The vertical solid lines marks the regions discussed in the text.}
\label{vel-prof}
\end{figure*}

The rotation curve and the radial velocity dispersion profile
along the major axis of NGC4778 are shown in Fig.\ref{vel-prof}.
The kinematic profiles reveal two important features: {\it i)} a
counter-rotation in the nuclear region, for $r\leq 2.5''$; {\it
ii)} at large galactocentric distances (for $r >2.5''$) the
rotation curve and velocity dispersion are asymmetric with respect
to the galaxy center. Note that the central value is chosen in
order to obtain the rotation curve symmetric within $3''$. Given
that, the dynamical center of the curves does not correspond to
the photometric one, the point of maximum signal to noise ratio,
which is offset by $2''$ in the NE direction.

On the whole, the galaxy presents significant rotation. An
increasing rotation is measured starting from $r\sim 3''$: on the
NE side, the rotation reaches a value of about 70 km/s at $r\sim
13''\ (\sim R_{e})$ and then it remain nearly constant out to
about $35''$\ ($\sim 2R_{e}$). At larger radii, as we have already
seen in the last section, the spectrum is contaminated by the
light coming from the galaxy NGC4761, therefore the value of $200\
km/s$ detected for $r = 55''$ cannot be due to the typical motion
of the stars in NGC4778. On the SW side, the velocity increases
reaching its maximum value of 80 km/s at $r\sim 20'' (\sim 1.4
R_{e})$; at larger radii we observe a drop in the velocity which
becomes consistent with zero at $r
> 35'' (\sim 2.5 R_{e})$.

The velocity dispersion decreases from the maximum value of about
$380 km/s$ in the center of the galaxy, to a value of $280-300\
km/s$ for $r > 5''$, remaining nearly constant, within the errors,
up to $2R_{e}$. Both on the SW and on the NE side, for $3''\leq r
\leq 5''$, the velocity dispersion decreases from 380 km/s to
280-300 km/s at $r \sim 5''$. At larger radii, the velocity
dispersion increases on the SW side, and reaches the maximum value
of about 420 km/s (at $r\sim 20''-25''$), and then decreases
reaching values consistent with those observed on the NE side.

Looking in more detail the nuclear region, for $r\leq 2.5''$ we
detect an inversion of the velocity gradient towards the center,
with a maximum value of the velocity of $20-30\ km/s$, with
respect to the outer radii. This inversion within the central
regions with respect to the overall trend, reveals the presence of
a counter-rotating decoupled core. In correspondence with the
inversion of the velocity gradient, the velocity dispersion
profile shows an hint for a local minimum (see Fig.\ref{vel-prof}
bottom panel).
\subsection{Photometry versus kinematics}\label{photkin}
\begin{figure*}
   \centering \includegraphics[width=8cm]{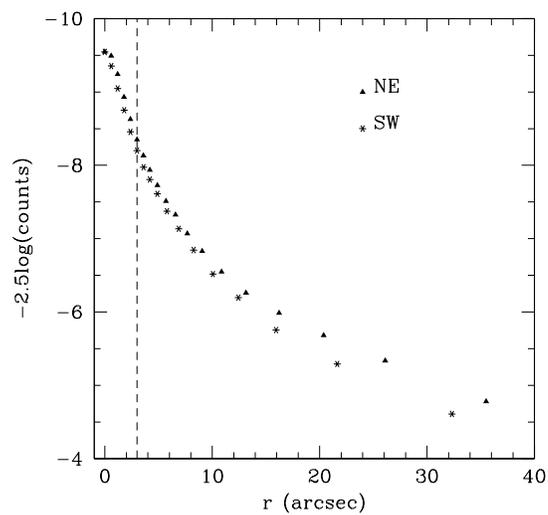}
   \caption{Comparison of the uncalibrated light profile along the
   slit in the NE and SW directions. The data-points were binned to
   reach a minimum signal-to-noise of 50.}  \label{prof-sn}
   \end{figure*}

In Fig.\ref{prof-sn} we plot the comparison of the light profile
along the slit in the NE and SW directions, where data-points were
binned to reach a minimum signal-to-noise of 50. For $r>3''$,
where the kinematic shows a clear asymmetry, the light profile
along the slit results unperturbed. This suggests that the
peculiarities in the outer kinematical properties may be
attributed to perturbed motion of the stars in NGC4778, rather
than to a contamination by the light from surrounding galaxies. In
particular, on the SW side, the increase detected at $r\sim 20''$
in both the rotation curve and the velocity dispersion profile, is
likely due to dynamical effects rather than to a projection
effect. Gravitational interactions, in fact, lead to the
distorsion of the orbits and a tidal heating of the stars
surrounding the interacting objects (cf. \cite{Combes95}).  For $r
\geq 12'' \sim R_e$, where the kinematic profiles on the SW side
also shows abrupt variations, we detect an increasing flattening
and twisting of the isophotes.

In order to check whether the observed flattening could be
attributed to the rotation or not, we evaluated the
\emph{anisotropy parameter} $(v/\sigma)^{\ast}$. The anisotropy
parameter $(v/\sigma)^{\ast}$ is defined as the ratio between the
observed value of $(v/\sigma)$ and the theoretical value for an
isotropic oblate rotator $(v/\sigma)_{t} =
[\epsilon/(1-\epsilon)^{1/2}]$ (\cite{Binney78}), where $\epsilon$
is the observed ellipticity. For NGC4778, at $r\sim 2 R_{e}$
($\sim 30''$), where $\epsilon \sim 0.4$, we estimate
$(v/\sigma)^{\ast}\sim 0.66$. Comparing this result with those
obtained by \cite{Bender94} for a sample of 44 elliptical
galaxies, we find that in the $M_{B} - log(v/\sigma)^{\ast}$
plane\footnote{$M_{B}$ is the total magnitude in the B band,
derived from the apparent magnitude within the isophote at $2
R_{e}$}, NGC4778 is located among the rotationally supported
galaxies (see Fig.18 of \cite{Bender94}).

In the nuclear regions, where kinematic shows a decoupled
component with an inversion of the velocity gradient, we have
$a_{4}/a > 0$, which indicates the presence of disky isophotes
(see Fig.\ref{iso} bottom panel).
%

\section{Discussion and conclusion}\label{discussion}

We have analyzed high signal-to-noise spectra along the major axis
of the dominant galaxy of the Hickson Compact Group HCG62. On the
whole, the observed kinematics and photometry of NGC4778 is
consistent with that of an S0 galaxy.

-\emph{Nuclear regions}- The higher resolution data enabled us to
detect the signature of the nucleus of NGC4778: inside $3''$ we
observe an inversion in the velocity profile gradient, which also
correspond to some anomalous photometric features, such as bluer
colors, and twisting in the position angle of the isophotes. These
features strongly suggest the existence of a small core ($\sim
600\ pc$) kinematically decoupled from the whole galaxy. Such
anomalous Kinematically Decoupled Cores (KDC) are common in
early-type galaxies and show very similar features to those
observed in the nuclear regions of HCG62a: the velocity profile is
characterized by a central asymmetry, to which corresponds an
unusual central isophotal flattening (see \cite{Krajnovic04}).

Similar behavior are observed for instance in NGC3623, belonging
to the Leo Triplet (\cite{Afanasiev05}). The central region of
this galaxy, in fact, shows the presence of a chemically distinct
core, a relic of a star formation burst, due to interactions, that
is shaped as a cold stellar disk with a radius of $\sim 250-350\
pc$. Like NGC4778, NGC3623 also shows a drop in the stellar
velocity dispersion in the nucleus. Numerical simulations
(\cite{Bournaud03}) predict that peculiar nuclear components may
be the result of an interaction event between two galaxies. Both
major merging and accretion of external material may induce that
some gas does flow in to the nuclear regions of the remnant and
quickly forms a small concentration of new stars that maintain the
original angular momentum of the initial galaxy, counter-rotate
with respect to the host galaxy. Furthermore, N-body simulations
including gas, stars and star formation, suggest that galaxies can
develope a central velocity dispersion drop due to nuclear gas
inflow, then subsequent star formation and the appearance of young
luminous stars born from dynamically cold gas (\cite{Wozniak06}).

-\emph{Formation and evolution}- Our data are in good agreement
with those obtained by \cite{Rampazzo98} along the direction
connecting the nuclei of NGC4778 and NGC4776 (P.A.$=128^\circ$).
They also found that the rotation curve is not symmetric with
respect to the center of NGC4778, with a rapid increase in the SE
direction at about $15''$ from the center. The velocity dispersion
increases on both sides, reaching a maximum at about $10''$ from
the center on the SW side. Given that the close galaxy NGC4776 is
located on the NW, they suggested that {\it i)} the rapid
variation of the rotation curve and the sharp increase of the
velocity dispersion in the SE direction is a real effect, reliable
due to a dynamical perturbation; {\it ii)} while the rise towards
NGC4776 (on the NW side) could be partly an artifact due to the
apparent superposition of the galaxies. The new kinematics along
the SW side (presented in this work) further suggest that the
South side of NGC4778 is dynamically perturbed.

As we have discussed in the section \ref{kin}, the rotation curve
of NGC4778 is not symmetric with respect to the center, and this
is a feature observed in many other compact groups
(\cite{Bonfanti99}). According to the literature, many compact
groups mainly composed by early-type galaxies, like HCG62, show
several morphological signature of interactions, and for all of
them the apparent kinematical interactions are not explainable as
a mere optical superposition. This conclusion is strongly
supported by the simulations performed by \cite{Combes95}, which
show that the peculiarities observed in many rotation curves of
galaxies belonging to compact groups are due to intrinsic effects
and not to contamination along the line of sight. Our results are
also in agreement with the estimates coming from merger
simulations (\cite{Combes95}), that predict asymmetry in the
kinematical profiles and a distinction between the photometric and
the dynamical center.

The asymmetry and the shape of the rotation curve and velocity
dispersion profile of NGC4778 do not find correlation with the
photometric features of the galaxy, except for the bluer colors in
the central region. The absence of correlation between the
dynamical and the morphological peculiarities suggests that the
dynamical properties of the HCG galaxies may be due to a minor
merger event. In fact, as showed by \cite{Nishiura00}, weak galaxy
collisions could not perturb the galaxy rotation curves, but
morphological deformations could be induced in the outer parts of
the galaxy (tidal tails, bridges etc), while minor mergers could
perturb the rotation curves in the inner regions, especially for
gas-poor early-type galaxies, without causing morphological
peculiarities.

We have estimated the mass-to-light (M/L) ratio of NGC4778 in
order to derive some constraints on the amount of dark matter in
HCG62. Since the kinematical profiles are not symmetric, for the
calculation of the M/L ratio, we have used the value of $v_{max}$
and $\sigma_{max}$ taken from the unperturbed side (NE) of the
curves. Moreover, in absence of an accurate photometric
calibration, we used as total B magnitude the value provided by
NED, $m_{b} = 13.79$. Choosing the values $v_{max}\simeq 80\
km/s$, $\sigma_{max}\simeq 270\ km/s$ and $R_{max}=34$ arcsec
($\simeq 10\ Kpc\ \simeq 2R_{e}$), by using the virial theorem
$(M/L)_{vir}\simeq\frac{2R\ (\sigma^{2}+v^{2})}{(L\ G)}$, we
obtain $M/L\simeq20.6$. This abnormally high mass-to-light ratio
is compatible with a recent merging which has induced a tidal
heating in the center of NGC4778, thus leading to a velocity
dispersion which is too high with respect to the actual mass of
the galaxy. This result however presents conflicting aspects. In
fact, while such behaviour is predicted by numerical simulations
(\cite{Combes95}), a detailed study of the x-ray diffuse halo
detected in the central regions of the group leads to a very
similar virial estimate of M/L. A more detailed discussion of this
point will be presented in \cite{Sodani06}.

The velocity dispersion of HCGs are generally higher than would be
expected given their visible mass (even if the discordant galaxies
are ignored): this can also be explained if the bulk of the mass
is in a non visible form (\cite{Hickson97}). Moreover, ROSAT
observations revealed a massive hydrogen envelope surrounding
HCG62, and showed that this group is dominated by dark matter.
Both N-body and hydrodynamic simulations indicate that the dark
matter halos of individual galaxies merge first, creating a
massive envelope within which the visible galaxies move (Barnes
1984, Bode et al 1993). Kinematic studies of loose groups (e.g.
Puche \& Carignan 1991) indicate that the dark matter is
concentrated around the individual optical galaxies. In contrast,
the X-ray observations indicate that in most compact groups, the
gas and dark matter are more extended and are decoupled from the
galaxies. This may be consistent with a M/L 30\% to 50\% lower in
compact groups respect to isolated galaxies (Rubin et al 1991).

The hierarchical mergers of cluster galaxies might power the
emission line gas in the center of the group members
(\cite{Valluri96}): according to the merger scenario, in order to
power emission-line nebulae, the merger must include a galaxy or a
group of galaxies that are late-types and which bring with them
cold gas. The observed $H_{\alpha}$ emission in NGC4778, and also
in NGC4776 and NGC4761, further support the idea that this galaxy
has recently experienced a merger event.

The overall scenario depicting NGC4778 as the product of a recent
merger, as emerges by previous discussion, is consistent with the
results obtained from X-ray observations. The presence of an
extended X-ray halo is consistent with scenarios describing
current compact group as the result of a first generation of
mergers, where the dominant galaxy sits at the bottom of a large
common gravitational well. The presence of two X-ray cavities in
the hot gaseous halo located on symmetrically with respect to
NGC4778 also indicate that the AGN residing in the galaxy core,
must have undergone a recent (a few $10^{7}$ yr, \cite{Birzan04})
active fase during which the radio-emitting relativistic plasma
has created two low density regions within the hot IGM. It is
commonly believed that such activity can be triggered by merging
events, which increases the accretion rate onto the central
massive black hole (\cite{Cattaneo05}). These fits a scenario in
which NGC4778 underwent a merger sometime in the past which
produced the counter-rotating core and triggered the nuclear
activity. However the low incidence of strong, type I AGN activity
in interacting galaxies suggests that a delay of several $10^8$
years is generally expected until the peak of the AGN fase
(\cite{Canal06, Grogin05}, however see also \cite{Koulouridis06}).
In the case of NGC4778 we can derive a lower limit of $10^{7}$ yr
for such delay from the age of the cavities. On the other end, an
upper limit is represented by the age of the merger which, given
the typical dynamical timescales of Compact Groups, can be
estimated in $~10^8$ yr. This result agrees with the estimate that
AGNs have duty cycles of the order of $10^{7-8}$ years.
\begin{acknowledgements}
The authors very grateful to R. Saglia for making available the
software for the kinematical study of early-type galaxies. M.S.
also wishes to thank the INAF-Observatory of Capodimonte for the
hospitality given during her thesis work, and N. Napolitano for
many useful discussions and suggestions. This work was funded
through a grant from Regione Campania (ex legge 5) and a MIUR
grant.  This work is based on observations made with ESO
Telescopes at the Paranal Observatories under programme ID
$<169.A-0595(C)>$ and $<169.A-0595(D)>$.

\end{acknowledgements}

\begin{thebibliography}{}

\bibitem[Afanasiev et al. 2005]{Afanasiev05} Afanasiev, V. L., and Sil'chenko, O.K., 2005, A\&A, 429, 825
\bibitem[Balcells et al. 1998]{Balcells98} Balcells, M., Gonzalez, C. G., 1998, ApJ, 505, 109
\bibitem[Bender 1990]{Bender90} Bender, R. 1990, A\&A, 229, 441
\bibitem[Bender et al. 1992]{Bender92} Bender, R., Burstein, D., and Faber, S. M., 1992, ApJ, 399, 462
\bibitem[Bender et al. 1994]{Bender94} Bender, R., Saglia, R.P., and Gerhard, O.E. 1994, \mnras, 269, 785
\bibitem[Bertola et al. 1999]{Bertola99} Bertola, F., and Corsini, E., 1999, Proceedings of IAU Symposium 186, "Galaxy Interactions at Low and High Redshift",
Edited by J. E. Barnes, and D. B. Sanders, p. 149
\bibitem[Bettoni et al. 1995]{Bettoni95} Bettoni, D., Buson, L. M., Maira, L. and Bertola, F., 1995, ASP conference series, 70, 95
\bibitem[Binney 1978]{Binney78} Binney, J., 1978, \mnras, 183, 501
\bibitem[Binney 1985]{Binney85} Binney, J., 1985, \mnras, 212, 767
\bibitem[Birzan et al. 2004]{Birzan04} Birzan L., Rafferty D. A., McNamara B. R. et al. 2004, ApJ, 607, 800
\bibitem[Bonfanti et al. 1999]{Bonfanti99} Bonfanti, P., Simien, F., Rampazzo, R. and Prugniel, Ph., 1999, A\&AS, 139, 483
\bibitem[Bournaud et al. 2003]{Bournaud03} Bournaud, F., and Combes, F., 2003, A\&A, 401, 817
\bibitem[Canalizo et al. 2006]{Canal06} Canalizo, G., Stockton, A., Brotherton, M. S. et al. 2006, Proceedings of the "QSO Host Galaxies: Evolution and Environment" conference, Lorentz Center, Universiteit Leiden, August 2005; 2006astro.ph..3218C
\bibitem[Caon et al. 1993]{Caon93} Caon, N., Capaccioli, M., and D'Onofrio, M., 1993, \mnras, 265, 1013
\bibitem[Cattaneo et al. 2005]{Cattaneo05} Cattaneo, A.; Combes, F.; Colombi, S. et al. 2005, MNRAS 359,1237
\bibitem[Combes et al. 1995]{Combes95} Combes, F., Rampazzo, R., Bonfanti, P.P., Prugniel, P. and Sulentic, J.W., 1995, A\&A, 297, 37
\bibitem[Coziol et al. 1998]{Coziol98} Coziol, R., Ribeiro, A. L. B., De Carvalho, R. and Capelato, H. V., 1998, ApJ, 493, 563
\bibitem[Eke 2004]{Eke04a} Eke, V.R., Frenk, C. S., Baugh, C. M., and co-authors, 2004, \mnras, 355, 769
\bibitem[Fukazawa et al. 2001]{Fukazawa01} Fukazawa, Y., Nakazawa, K., Isobe, N., Ohashi, T. and Kamae, T., 2001, ApJ, 546, 87
\bibitem[Fukugita et al. 1995]{Fukugita95} Fukugita, M., Shimasaku, K., Ichikawa T., 1995, Publications of the astronomical society of the Pacific, 107, 945
\bibitem[Gerhard 1993]{Gerhard93} Gerhard, O. E., 1993, \mnras, 265, 213
\bibitem[Governato et al. 1991]{Governato91} Governato, F., Bhatia, R. and Chincarini, G., 1991, ApJ, 371, 15
\bibitem[Grogin et al. 2005]{Grogin05} Grogin, N. A., Conselice, C. J., Chatzichristou, E., et al. 2005, ApJ, 627L, 97
\bibitem[Hickson 1997]{Hickson97} Hickson, P., 1997, Annual review of Astronomy and Astrophysics, 35, 357
\bibitem[Iodice et al. 2001]{Iodice01} Iodice, E., D'Onofrio, M., Capaccioli, M., 2001, Astrophysics and Space Science, 276, 869
\bibitem[Jog et al. 2006]{Jog06} Jog, C. J., and Maybhate, A., 2006, \mnras, 000, 000
\bibitem[Koulouridis et al.2006]{Koulouridis06} Koulouridis, E., Plionis, M., Chavushyan, V., Dultzin-Hacyan, D., Krongold, Y., \& Goudis, C.\ 2006, \apj, 639, 37
\bibitem[Krajnovic et al. 2004]{Krajnovic04} Krajnovic, D., and Jaffe, W., 2004, A\&A, 428, 877
\bibitem[Kropolin et al. 2000]{Kropolin00} Kropolin, W., and Zeilinger, W. W., 2000, A\&AS, 145, 71
\bibitem[Mehlert et al. 2000]{Mehlert00} Mehlert, D., Saglia, R., Bender, R., and Wegner, G., 2000, A\&AS, 141, 449
\bibitem[Mendes de Oliveira 1992]{Mendes92} Mendes de Oliveira, C., 1992, PhD Thesis, Univ. of British Columbia
\bibitem[Mendes de Oliveira et al. 2003]{Mendes03} Mendes de Oliveira C., Aram, P., Plana, H. and Balkowski, C., 2003, The astronomical journal, 126, 2635
\bibitem[Nishiura et al. 2000]{Nishiura00} Nishiura, S., Shimada, M., Ohyama, Y., Murayama, T. and Taniguchi, Y., 2000, The astronomical journal, 120, 1691
\bibitem[Pildis et al. 1996]{Pildis96} Pildis, R. A., Evrard, A. E., and Bergman, J. N., 1996, The astronomical journal, 112, 378
\bibitem[Rampazzo et al. 1998]{Rampazzo98} Rampazzo, R., Covino, S., Trinchieri, G., and Reduzzi, L. 1998, A\&A, 330, 423
\bibitem[Rubin et al. 1991]{Rubin91} Rubin V.C., Hunter D.A., Ford W.K.Jr. 1991, ApJS, 76, 153
\bibitem[Sodani et al. 2006]{Sodani06} Sodani S., De Filippis E., Paolillo M., Longo G., Spavone M., 2006, in preparation (Paper II)
\bibitem[Tully 1987]{Tully87} Tully, R. B. and Fisher, J. R., 1987, \skytel, 74, 612
\bibitem[Valluri et al. 1996]{Valluri96} Valluri, M. and Anupama, G.C., 1996, The astronomical journal, 112, 1390
\bibitem[Vennik et al. 1993]{Vennik93} Vennik, J., Richter, G. M. and Longo, G., 1993, AN, 314, 393
\bibitem[Vrtilek et al. 2002]{Vrtilek02} Vrtilek, J., M., Grego, L., David, L. P. et al., 2002, APS meeting, B17.107
\bibitem[Wozniak et al. 2006]{Wozniak06} Wozniak, H. and Champavert, N., 2006, \mnras, 000, 000
\bibitem[Zabludoff \& Mulchaey 1998]{Zabludoff98} Zabludoff, A. I. and Mulchaey, J. S., 1998, ApJ, 498, L5
\bibitem[Zepf et al. 1991]{Zepf91} Zepf, S. E., Whitmore, B. C., Levison, H. F., 1991, ApJ, 383, 524

\end{thebibliography}
\end{document}